Article

# Supernovae, Neutrinos, and the Chirality of the Amino Acids

R.N. Boyd[1]*, T. Kajino[2,3], and T. Onaka[3]

[1]Lawrence Livermore National Laboratory
[2]National Astronomical Observatory of Japan
[3]University of Tokyo, Department of Astronomy, Graduate School of Science

*Author to whom correspondence should be directed:
Boyd11@llnl.gov
925-423-3201
FAX: 925-292-5940

**Abstract:** A mechanism for creating an enantioenrichment in the amino acids, the building blocks of the proteins, that involves global selection of one handedness by interactions between the amino acids and neutrinos from core-collapse supernovae is described. The chiral selection involves the dependence of the interaction cross sections on the orientations of the spins of the neutrinos and the $^{14}$N nuclei in the amino acids, or in precursor molecules, which in turn couple to the molecular chirality. It also requires an asymmetric distribution of neutrinos emitted from the supernova. The subsequent chemical evolution and galactic mixing would ultimately populate the Galaxy with the selected species. The resulting amino acids could either be the source thereof on Earth, or could have triggered the chirality that was ultimately achieved for Earth's proteinaceous amino acids.

## 1. Introduction

A longstanding puzzle in biology and astrobiology has been the existence of left-handed amino acids and the virtual exclusion of their right-handed forms [1-4].  This is especially puzzling because most mechanisms suggested for creating this enantiomerism would create one form in nature locally but would create equal numbers of the other somewhere else.  The total dominance of the left-handed forms on Earth is well known, but the left-handed forms appear also to be preferred beyond the locality of Earth, based on meteoritic evidence [5-8].  This is especially interesting for two reasons: in the absence of data to the contrary it must be recognized that the preference of the left-handed forms may persist throughout the cosmos, but both forms were apparently frozen into the meteorites before the right-handed forms could be eliminated. However, another possibility is that the meteoroids (which become meteorites when they strike

Earth) were strongly enantiomeric, but became racemized in their journey through space, or on their passage through Earth's atmosphere [9-11]. Recent work [12] has confirmed the preference for the left-handed forms.

It is generally accepted that if some mechanism can introduce an imbalance in the populations of the left- and right-handed forms of any amino acid [13], successive synthesis or evolution of the molecules involving autocatalytic reactions and out-of-equilibrium thermochemistry amplify this enantioenrichment to produce ultimately a single form. What is not well understood, though, is the mechanism by which the initial imbalance can be produced, and the means by which it always produces the left-handed chirality for the amino acids (if, indeed, that is the case). The energy states of the left- and right-handed forms have been shown, by detailed computations, to differ at most by infinitesimal amounts [14, 15], so it would be difficult for thermal equilibrium to produce the imbalance.

One suggested mechanism lies with processing of a population of amino acids, or of their chiral precursors, by circularly polarized light [18-23]; this could select one chirality over the other. However, this solution does not easily explain why it would select the same chirality in every situation, as is apparently observed (albeit with limited statistics), or why the physical conditions that would select one form in one place would not select the other in a different location. One possibility [20], that a region as small as a planetary system could be processed by the output from a localized region of a single star so that all of the light could be of a single circular polarization, could explain the observed (local) meteoritic and Earth's results, although it would have to be assumed that mixing subsequent to enantioselection would not mitigate the resulting enantiomerism to too great an extent. In addition, the circularly polarized light model must photolyze, that is, destroy, large amounts of amino acids in order to produce a significant enantioenrichment. Another possibility [24] invoked selective processing by some manifestation of the weak interaction, which does violate parity conservation, so might perform a selective processing. This idea was based on earlier work [25, 26]. Ref. [24] focused on the β-decay of $^{14}$C to produce the selective processing. However, it was not possible in that study to show how simple β-decay could produce chiral-selective molecular destruction. A modern update on this possibility [27] appears to produce some enantioenrichment. Another suggestion [28] assumed that neutrinos emitted by a core-collapse supernova would selectively process the carbon or the hydrogen in the amino acids to produce enantiomerism. This suggestion also did not explain how a predisposition toward one or the other molecular chirality could evolve from the neutrino interactions. A similar suggestion [29] involves the effects of neutrinos from supernovae on molecular electrons. Another suggestion involved the differences between ortho and para hydrogen pairs [30] in the amino acids, which could produce an effect on them.

In this paper we also invoke the weak interaction to perform selective destruction of one molecular chirality. This expands on a previously published paper [31]. The key to this mechanism is the selective processing of the molecules that have been observed to exist [32] in molecular clouds in the vicinity of a core-collapse supernova, but would be specific to $^{14}$N, a nucleus with spin of 1 (in units of Planck's constant, $\hbar$) that exists in all amino acids. Two features of supernovae are important: the magnetic field that is established as the star collapses to a neutron star or a black hole and the intense flux of neutrinos that is emitted as the star cools. The magnetic fields only have to couple to the molecules strongly enough to produce some orientation of their non-spin-zero nuclei. However, the $^{14}$N must also couple in some way to the molecular chirality. In the model we assume to describe this the neutrinos preferentially interact with the $^{14}$N atoms in one of the chiral forms, and convert the $^{14}$N to $^{14}$C, thereby destroying that molecule and so preferentially selecting the other chiral form. This is discussed in Section 2.

Another question involves the extent to which a single chirality might populate the entire Galaxy [33]. Although supernovae could not do so by themselves, subsequent chemical amplification of the chirality-selected, biologically-interesting molecules would amplify the enantiomerism of the dominant form. Then Galactic mixing, operating on a slower timescale, would be able to establish the dominant form throughout the Galaxy. These two mechanisms would make it likely that the Galaxy would be populated everywhere with the same preferred chiral form. These features are described in Section 3.

In Section 4 we address two potential issues for this model, one being the possibility that other nuclei in the amino acids might produce similar effects to those from $^{14}$N, and the other being constraints that the radiation from the supernovae that produce neutrinos that perform the chiral selection, or from their progenitor stars, which in neither case is expected to be polarized, might impose on the parameters of the model.

Finally, Section 5 presents our conclusions.

## 2. Selective Destruction Mechanism

The key to selective processing of this model is the $^{14}$N nucleus. Although the spatial arrangement of the molecular electrons determines the chirality of the molecule, the $^{14}$N nuclear spin would couple to the electronic spin, which would couple to an external magnetic field, thereby aligning the spin 1 nuclei to that magnetic field to some extent, and providing an orientation direction of the $^{14}$N spin in the molecules. Also important is the chirality of the (spin ½) neutrinos: electron antineutrinos, $\underline{\nu}_e$s, are right-handed and electron neutrinos, $\nu_e$s, are left-handed [34]. When the spin of the $^{14}$N is aligned with that of the electron antineutrino, from simple angular momentum coupling, the total neutrino-nuclear spin would be 3/2, but if the two spins were antialigned, there would be a mixed ½ and 3/2 total spin, and a different magnetic substate distribution. The reaction

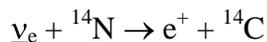

$$\underline{\nu}_e + {}^{14}\text{N} \to e^+ + {}^{14}\text{C}$$

would not have the same strength for these two spin configurations, as discussed below, so the resulting destruction of the $^{14}$N, and hence of the molecule, would depend on the electronic structure, hence the chirality or symmetry, of the molecule. The above (nuclear) transition, in the nomenclature of standard, and experimentally well documented, beta-decay, is pure "Gamow-Teller" [34, 35], since it is between nuclei with spins that differ by one unit (this also applies to the $\nu_e+{}^{14}$N $\to$ e$^-+{}^{14}$O reaction, discussed below). The geometry of this situation is illustrated in figure 1.

The Q-values of the reactions are critical to this argument; that for $\underline{\nu}_e+{}^{14}$N $\to$ e$^+ +{}^{14}$C is -1.18 MeV, and for $\nu_e+{}^{14}$N $\to$ e$^-+{}^{14}$O is -5.14 MeV. The high threshold energy for the latter reaction would inhibit it compared to the former, as the energies of the neutrinos emitted in the stellar collapse are comparable to those Q-values [36-40]. In addition, the cross sections for the neutrino induced reactions increase as the square of the energy above threshold [34]. Thus while many electron neutrinos would be able to convert $^{14}$N to $^{14}$O, many more electron antineutrinos could drive $^{14}$N to $^{14}$C producing a chiral preference by preferentially destroying the molecules with the other chirality [31]. Any additional (second order) effects produced by the electrons or positrons [29] resulting from the neutrino-nucleus interactions would be expected to be completely negligible.

The magnitude of the selection in the extreme situations for $\underline{\nu}_e + {}^{14}N \rightarrow e^+ + {}^{14}C$, i.e., those in which the antineutrino and $^{14}N$ spins are either aligned or antialigned, can be estimated as follows. If the spins are aligned, the angular momentum of the combined $^{14}N$(spin 1) + $\underline{\nu}_e$(1/2) system must be 3/2, whereas if they are antialigned it will be ½. The transition produces $^{14}C(0)$. Thus, since the weak interaction can be assumed to be point-like for the present situation, conservation of angular momentum in the aligned case requires some orbital angular momentum transfer. This can be $\ell=1$ for the weak interaction (again, in units of Planck's constant, $\hbar$), but this will inhibit the transition compared to that for the antialigned case, in which no orbital angular momentum transfer is required, by roughly an order of magnitude [34, 35]. Since the angular momentum coupling coefficients favor the case in which the spins of the $^{14}N$ and $\underline{\nu}_e$ are antialigned by 2:1 over that in which they are aligned, the antialigned $^{14}N$ nuclei will suffer selective destruction by roughly a factor of seven (two-thirds of the order of magnitude) over the aligned nuclei.

Although determination of the total effect on the selective destruction of the chiral molecules would require integration over all space, the average effect may not be relevant. If amplification of the enantiomerism produced by any effect progresses more rapidly where the enantiomeric excess is larger, the relevant number will most likely be the maximum enantiomeric excess achieved. We return to this point in section 3.

**Figure 1.** Schematic diagram of the magnetic fields, $\underline{B}$, surrounding a nascent neutron star and the spins of the $^{14}N$ nuclei, $S_N$, and of the neutrinos, $S_\nu$, emitted from the supernova that created the neutron star.

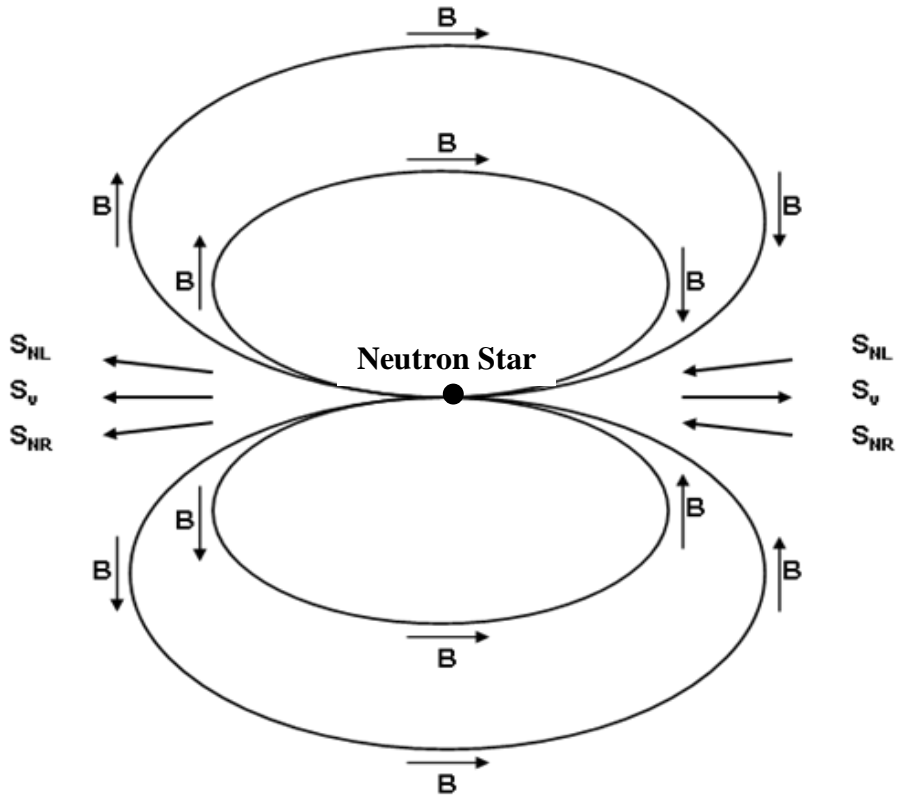

To illustrate how the $^{14}$N spin might couple to the chirality of the molecules of which they are a part, we have assumed a model that was developed by Buckingham [41, 42] in the context of nuclear magnetic resonance. Paraphrasing from Ref. [41], the magnetic field induces a current density which, in chiral samples, causes a nuclear magnetic moment (odd under time reversal and even parity) to induce a molecular electric dipole moment (even under time reversal and odd parity). The result is an effect that is opposite in sign for left-handed and right-handed enantiomers. The critical feature for this model [41] is the necessity of having a non-zero spin nucleus, i.e., the $^{14}$N, for this effect to occur at all.

The situation that exists is illustrated in fig. 2. The result of the Buckingham effect is to skew the populations of the substates, as is indicated, with the skewing going in opposite directions for the LH and RH molecules. Although the magnetic field interacts with the atomic angular momenta, the nuclear spin will tend in the same direction as that of the atomic angular momentum. The resulting LH and RH populations will be the same at either throat of the neutron star, but a combination of the interactions with the electron antineutrinos and an asymmetry in the neutrino flux at the two throats will perform a chiral selection. The magnetic substates that undergo greater processing are indicated by an "x" in fig. 2. As can be seen, at the magnetic-field-outgoing throat, the +1 magnetic substate of the NL molecules will be selected, with the other two substates undergoing relatively greater destruction from the antineutrino interactions. The -1 magnetic substate of the NR molecules will be preferentially selected by the Buckingham effect, and it will undergo selective destruction. This will create a preference for the NL molecules over the NR molecules at that throat of the neutron star. Technically we know only that the Buckingham effect will drive the NR and NL molecules in opposite directions, but we cannot be certain which molecular chirality it will favor without studying specific molecules.

**Figure 2.** Magnetic substate distributions for $^{14}$N nuclei resulting from the Buckingham effect. The "x" indicate the states that will undergo greater processing from the electron antineutrinos. NL refers to left-handed $^{14}$N nuclei, and NR to the right-handed $^{14}$N nuclei. "Outgoing (Incoming) throat" refers to the throat of the neutron star at which the magnetic field is outgoing (incoming). The thickness of the line for each substate indicates its relative population.

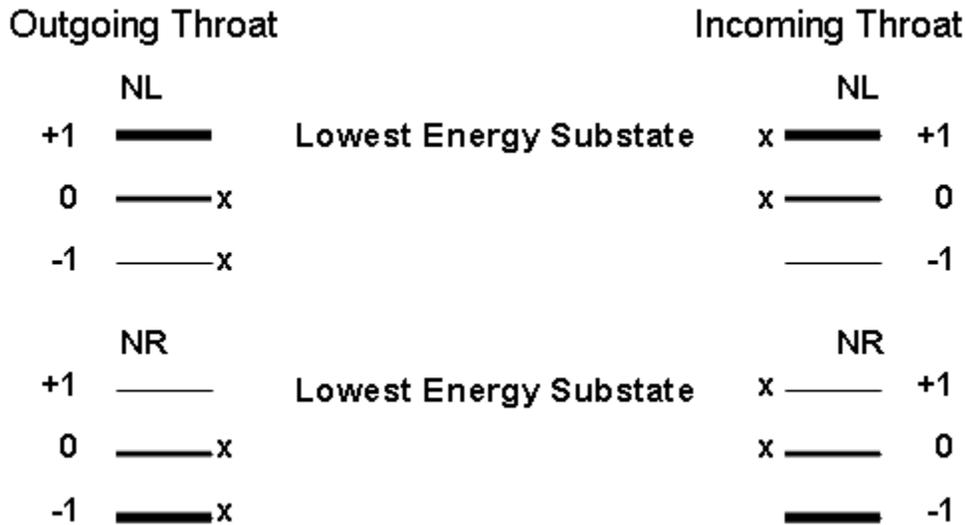

At the magnetic-field-incoming throat, the exact opposite selection effect occurs, and if the neutrino fluxes were the same at the two throats, no net chiral selection would occur. However, recent work [43-46] has shown that the extraordinary magnetic field of the neutron star will modify the cross sections for neutrino capture, decreasing them by 20-30 percent near zero degrees on the magnetic-field-outgoing throat and increasing them by a few percent at the opposite throat. *This will produce a neutrino flux asymmetry, hence a greater selection of one, if the model is correct, left-handed, chirality molecules at the magnetic-field-outgoing throat, and a net overall chiral selection [47] that will be the same for every neutron star or black hole.* Note, though, that selection of right-handed molecules would occur at the magnetic-field-incoming throat, and the large spatial separation of the two regions might allow for some isolated grains to contain right-handed molecules, even though the overall selection would be toward left-handed molecules. If mixing is sufficiently thorough, however, only the single chirality will ultimately prevail.

**3. Spreading Enantiomerism Throughout the Galaxy**

To what extent would all the molecules in the Galaxy be processed by the effect being described? The supernovae by themselves do not come close to being able to do so [31]. We believe that a more plausible mechanism for driving the entire Galaxy to some level of enantiomerism is a combination of chemical amplification and Galactic mixing. As soon as a supernova explodes the enantiomeric material it has produced begins to mix with the racemic material adjacent to its processed volume, driving the racemic material with which it mixes toward partial enantioenrichment, possibly on the surfaces of the dust grains [48] that also exist in the molecular clouds. This ultimately would extend the enantioenriched volume well beyond that which was processed by the supernova neutrinos.

A collection of molecules exhibiting a small enantiomeric excess can amplify that excess dramatically as has been demonstrated for some environments. A general discussion of this capability [50] assumed that such enantiomeric excess might be produced by weak neutral currents, even from fluctuations [49], and then amplified toward much greater enantioenrichment by chemical replication. The chemical replication of molecule X would proceed [49, 51] by autocatalysis, and would preserve chirality, for example, by S+T↔X, which then has its chirality established so to become $X_L$. Replication and amplification then could occur by S+T+$X_L$↔2$X_L$. The ability of autocatalysis to produce amplification of enatiomerism has been demonstrated in the laboratory [52, 53]. A subsequent study [54] confirmed the possibility of autocatalysis, and did so in an environment that was relevant to amino acid autocatalysis. Ref. [55] showed that the amino acid enantiomerism could be amplified by successive evaporations to precipitate the racemate, with the solution becoming highly enantioenriched. Although this would not be relevant to amplification in dust grains, it could produce additional amplification once the somewhat enantiomeric grains or meteorites landed in a suitable planetary environment.

More specific to cosmic amplification, a model was developed [56] in which chiral replication of complex molecules would occur in the interstellar medium in the warmed (possibly by the proximity to a supernova) ice outer shells of grains. In that model, chemical replication was catalyzed by radicals, for example, H, OH, CO, $CH_3$, NH, and $NH_2$, created by the interactions of high energy cosmic rays with preexisting molecules. Subsequent studies, both theoretical [57, 58] and experimental [59] suggest that the conditions of outer space will permit the chemistry necessary to produce a variety of complex molecules. Thus it appears that either autocatalysis or chemical replication via radicals, or other complex chemical processes, could produce the amino acids and the chemical replication required for the present model to succeed, and would extend the chirality established by the magnetic fields and neutrinos from supernovae.

As the enantiomeric excess of the processed material increased via chemical replication, it will also mix with the rest of the galactic material, ultimately establishing a preference for left-handed amino acids throughout the Galaxy. Details of the processes by which this would occur have been discussed [60], and include many types of astronomical sources. The galactic mixing timescale is much smaller than the age of the universe [61], so the enantioenrichment established locally would be shared throughout the Galaxy. As one signature of galactic mixing time, our Galaxy rotates roughly once every $3 \times 10^8$ years [62], much less than the $\sim 12 \times 10^9$ years the Galaxy has lived. The evolutionary timescales of organic molecules are undoubtedly much shorter than the galactic mixing timescales. Although this might depend on many variables, the fact that such molecules are born in the molecular clouds, and that these clouds are born, live, and die in of order 10 My [63, 64], confirms the shortness of the chemical evolutionary timescale.

This model for propagating the supernova-selected enantiomerism throughout the Galaxy allows for the same chirality to exist on a large scale. However, as noted in section 2, some right-handed chiral molecules will also exist locally. The numbers of these right handed molecules would be appreciably less than those of the left-handed ones, but they might be observed in, for example, the meteorites in which enantiomerism, but not homochirality, is found to exist. One new test may occur before long; it is hoped that the Japanese satellite Hayabusa [65-67], which returned to Earth in 2010, may have some dust samples from asteroid Itokawa. If our model is correct, most of the inclusions of amino acids in those samples must also have chiralities consistent with those of their terrestrial counterparts. An additional test will occur in 2014 when the ROSETTA mission sends a lander onto Comet 67 P/Churyumov-Gerasimenko [68].

## 4. Potential issues for this model

Several questions immediately arise. One involves $^2H$, another spin 1 nucleus, and a $1.5 \times 10^{-4}$ component of hydrogen. $^1H$ might have a similar effect. These issues are discussed in [31]; they are found not to be capable of producing a significant effect.

A more serious concern with this model is with the extraordinary photon field from the supernova to which the molecules, and the grains on which they are thought to form [69, 70], would be subjected. The solution to this problem lies with the supernovae that result in collapse to a black hole. Recent studies [71, 72] suggest that stars having masses from 8 to 25 times that of the Sun would be expected to form a neutron star, stars from 25 to 40 solar masses would form a neutron star, but the fallback would ultimately produce a black hole, and stars having masses more than 40 solar masses would collapse directly to a black hole. Ref. [71] also showed that roughly the same number of neutrinos would be emitted from stars in the latter mass range as for those that collapse to a neutron star, but that a larger fraction of the neutrinos would be electron antineutrinos and that their energies would be higher than in the supernovae from less massive progenitor stars. Both of these effects would enhance the chiral processing. For stars with masses greater than 40 solar masses, the result would be a "failed supernova", i.e., the photon flux would be small or nonexistent. The stars in the 25 to 40 solar mass range would produce some photons, but they would be suppressed compared to the neutron-star case. They might produce even more neutrinos than the stars that produce neutron stars directly [73].

Thus the supernovae that go directly to black holes would both produce the neutrinos required to produce molecular enantioenrichment and not then destroy the molecules with photons. Because the time required for fallback black hole formation is short, many of these black hole forming supernovae [74] might also be expected to produce too few photons to destroy the molecules that their neutrinos had processed. Supernovae that produce a neutron star are more frequent than those that produce a black hole [75], but the fraction that does produce a black hole is significant (~20% [71]).

Another potential concern is with the progenitor star. A red giant would encompass the entire region that the supernova could process; these would be the progenitors of the 8-25 solar mass stars. Wolf-Rayet stars, thought to be the progenitors of the supernovae resulting from stars of more than 25 solar masses, are small, but are very hot. However, meteoroids that passed through the Wolf-Rayet clouds would be exposed to the highest temperature regions of the clouds through which they passed for only short periods of time. These would have been formed elsewhere, and could contain molecules formed as they passed through their giant molecular cloud. And there would be a large number of these grains and meteoroids that could produce enantiomerism. They only have to approach within ~one AU of the Wolf-Rayet star to be processed, and this is certainly a large enough volume to include a huge number of grains and meteoroids.

Although molecules on the grain or meteoroid surfaces would certainly get evaporated or dissociated once they reach the vicinity of the star, if they consisted of agglomerated grains, the molecules that resided on the surfaces of the internal grains could be retained. Such agglomerations would have to be large enough initially to withstand some radiative ablation as they passed through the Wolf-Rayet cloud, with the extent of the surface ablation depending on their closeness of approach to the central star. None the less, these grains and meteoroids appear to provide the best opportunity for relatively highly enantioenriched samples to be produced. Using estimated neutrino-nucleus cross sections [76], we estimate that an object passing within a distance of $10^{12}$ cm of the star would achieve a processing probability of roughly $10^{-9}$, and therefore, an enantiomeric excess of $0.5 \times 10^{-9}$. If it were large enough to survive if it passed by at $10^{11}$ cm, it would achieve a maximum enantiomeric excess of roughly $5 \times 10^{-6}$ percent, or possibly even higher, given the possible enhancements of electron antineutrinos from the massive supernovae.

## 5. Conclusions

If this model turns out to be correct, the longstanding question of how the organic molecules necessary to create and sustain life on Earth were created will have undergone a strong suggestion that the processes of the cosmos played a major role in establishing the molecules of life on Earth, either directly, or by providing the seeds that ultimately produced homochirality in the amino acids. These molecules would appear to have been created in the molecular clouds of the galaxy, with their enantiomerism determined by supernovae, and subsequently either transported to Earth only in meteorites, swept up as the Earth passed through molecular clouds, or included in the mixture that formed Earth when the planets were created. Any scenario in which these molecules were created exclusively on Earth in Darwin's "warm little pond," and supported by the experiment of Ref. [77], would find it much more difficult to explain the enantiomerism that is observed on Earth and, apparently, generally in the cosmos.

**Acknowledgements**

This work has been supported by the US National Science Foundation grant PHY-9901241, and under the auspices of the Lawrence Livermore National Security, LLC, (LLNS) under Contract No. DE-AC52-07NA27344. This paper is LLNL-JRNL-474633. The authors express their gratitude for helpful interactions with N. Sleep, I. Tanihata, R. Kuroda, L.D. Barron, L. Fried, and E. Branscomb.